\documentclass[conference, 10pt]{IEEEtran}
\usepackage{cite,graphicx,amsmath,psfrag,bm}
\usepackage[tight,footnotesize]{subfigure}
\usepackage[usenames,dvipsnames]{xcolor}
\usepackage{mathabx}
\usepackage{cancel}
\usepackage{epstopdf}
\usepackage{stfloats}
\usepackage{wrapfig}
\usepackage[outercaption]{sidecap}
\usepackage{pstool}
\usepackage[justification=centering]{caption}
\usepackage{float}

\title{Diagrammatic Explanation of the Reverse Doppler Effect in Space-Time Modulated Photonic Crystals}

\author{\IEEEauthorblockN{Zo\'e-Lise Deck-L\'eger\IEEEauthorrefmark{1},
Maksim Skorobogatiy\IEEEauthorrefmark{2}
and
Christophe Caloz\IEEEauthorrefmark{1}}
\IEEEauthorblockA{\IEEEauthorrefmark{1}Dpt. of Electrical Engineering,
Polytechnique Montr\'eal, Montr\'eal, QC H3T 1J4, Canada\\ Email: see http://www.calozgroup.org}
\IEEEauthorblockA{\IEEEauthorrefmark{2}Dpt. of Physics Engineering,
Polytechnique Montr\'eal, Montr\'eal, QC H3T 1J4, Canada \\
}}

\date{\today}
\begin{document}
\maketitle
\newcommand{\ud}{\,\mathrm{d}}
\newcommand{\D}{\,\partial}
\newcommand{\rect}{\,\mathrm{rect}}
\newcommand{\sinc}{\,\mathrm{sinc}}

\begin{abstract}An inverse Doppler shift occurs in a photonic crystal (PC) bounded by a moving wall. The interpretation of this result has stirred some controversy. In this paper, we address the problem using a diagrammatic approach. This visual representation provides immediate insight into the phenomenon, and is a powerful tool for the design of time-varying PCs.
\end{abstract}

\section{Introduction}

\begin{figure}[h]
\centering
\psfragfig*[width=0.4\columnwidth]{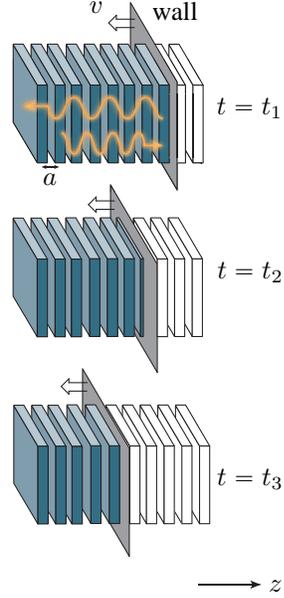}{
\psfrag{z}[c][c]{$z$}
\psfrag{1}[c][c]{$t=t_1$}
\psfrag{2}[c][c]{$t=t_2$}
\psfrag{3}[c][c]{$t=t_3$}
\psfrag{v}[c][c]{$v$}
\psfrag{P}[l][l]{wall}
\psfrag{a}[c][b]{$a$}
}
\centering
\caption{Semi-infinite photonic crystal (PC) of period $a$ bounded by a mirror moving in the negative $z$ direction with velocity $v<0$ , at three time instants. The incident wave is propagating in the positive $z$ direction.}
\label{fig:ch5_3D_geometry}
\end{figure}

There has been a recent interest for space-time \mbox{(S-T)} systems. S-T systems simultaneously transform the spatial and temporal spectra of electromagnetic waves, which gives rise to a variety of novel effects and potential applications. In this paper, we study the S-T system depicted in Fig.~\ref{fig:ch5_3D_geometry}, which consists of a photonic crystal (PC) bounded by a moving mirror. The wall can be experimentally implemented by modulating a PC with a shock wave as in~\cite{reed2003}. This structure has been shown to generate harmonics obeying the inverse Doppler effect~\cite{reed2006patent,reed2003,seddon}, and has generated some controversy regarding the physical interpretation of this phenomenon~\cite{seddoncomment,reedcomment}. We address here the problem using a diagrammatic approach inspired by works done in the 1960's on periodic space-time systems\cite{olinercassedy1963,tamirwang1966,chutamir1972,chutamir1969}.

\section{PC Bounded by a Stationnary Wall}

We first study the case of a PC bounded by a stationary wall to introduce the approach and the terminology.

\begin{SCfigure*}
\centering
\psfragfig*[width=1.4\columnwidth]{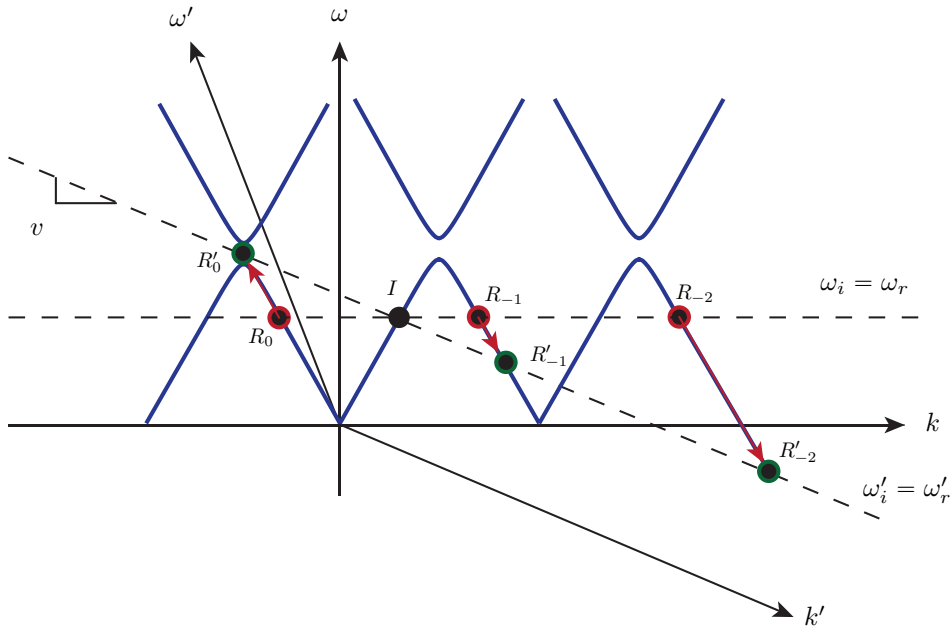}{
\psfrag{k}[c][c]{$k$}
\psfrag{w}[c][c]{$\omega$}
\psfrag{a}[c][c][0.8]{$R_0$}
\psfrag{b}[c][c][0.8]{$R_{-1}$}
\psfrag{c}[c][c][0.8]{$R_{-2}$}
\psfrag{d}[c][c][0.8]{$R'_0$}
\psfrag{e}[l][c][0.8]{$R'_{-1}$}
\psfrag{f}[c][c][0.8]{$R'_{-2}$}
\psfrag{I}[c][c][0.8]{$I$}
\psfrag{i}[c][c]{$\omega_i=\omega_r$}
\psfrag{j}[l][c]{$\omega'_i=\omega'_r$}
\psfrag{K}[c][c]{$k'$}
\psfrag{W}[c][c]{$\omega'$}
\psfrag{v}[c][c]{$v$}
}
\captionsetup{justification=justified}
\caption{PC dispersion diagram, in blue, with superposed ($\omega, k$) and Lorentz transformed ($\omega', k'$) axes. Incident, $I$, wave is on fundamental harmonic. Reflected harmonics are at intersections with isofrequency lines and are indicated by $R$, in red, for the stationary wall and $R'$, in green, for the moving wall. We notice an inverse Doppler shift for the $R'_{-1}$,$R'_{-2}$ solutions. The slope of the isofrequency curve in the moving frame is the velocity of the wall $v$}
\label{fig:ch5_dispersion}
\end{SCfigure*}

The dispersion equation for a periodic structure with an infinitesimally small modulation of the refractive index, \mbox{$\Delta n \rightarrow 0$}, is given by
\begin{equation}
(k+2m\pi/a)^2=(\omega n/c)^2,~m=\ldots,-2,-1,0,1,2,\ldots
\end{equation}
\noindent where $a$ is the period of the PC. As $\Delta n $ increases, the corresponding dispersion curves differ from the infinitesimally small modulation curves essentially only at their intersections, where stopbands opens up as a result of Bragg diffraction. The dispersion diagram for such a periodic medium is shown in Fig.~\ref{fig:ch5_dispersion}. The incident wave is concentrated on the fundamental, $n=0$, space harmonic in the figure. Since the wall is stationary, the temporal frequency of the reflected wave is unchanged, $\omega_r=\omega_i$. Therefore, the solutions lie on the dashed horizontal line. For small modulations of the refractive index, it is sufficient to consider only the fundamental ($n=0$) and next few harmonics ($n=\pm1,\pm2$)~\cite{tamirwang1966}. We only represent the first negative harmonics in the diagram, because these exhibit interesting physics, as will be shown next.

\section{Photonic Crystal Bounded by a Moving Wall}

We now want to solve the case where the wall is moving. The medium has not changed, and so the same dispersion relation holds. However, the boundary condition at the wall must be treated differently. For this purpose, we superimpose on the previously described dispersion diagram the axes corresponding to a frame of reference moving with the wall, using the Lorentz transform for spatial and temporal frequencies, as
\begin{subequations}
\begin{equation}
\omega'=\frac{1}{\sqrt{1-(\frac{v}{c})^2}}\left(\omega-v k\right),
\end{equation}
\begin{equation}
k'=\frac{1}{\sqrt{1-(\frac{v}{c})^2}}\left(k-\frac{v}{c}\frac{\omega}{c}\right).
\end{equation}
\end{subequations}

The $k'$ axis corresponds to $\omega'=0$, or $\omega=v k$ while the $\omega'$ axis corresponds to $k'=0$, so $k=v/c^2 \omega$, where $v<0$ consistently with Fig.~\ref{fig:ch5_3D_geometry}. One next draws an isofrequency line parallel to the $k'$ axis. This line intersects the dispersion diagram at a series of new points on the positive and negative space harmonics. Projecting this points to the rest-frame frequencies ($\omega$) reveals that the negative harmonics ($n=-1,-2$), on the right side, have been down-shifted, whereas the fundamental $n=0$ reflection harmonic has been up-shifted. This means that an inverse Doppler shift effect has occurred in the negative harmonics.

From this point, the frequencies of the reflected field are readily found from basic geometry as
\begin{equation}
\omega_{\text{ref},m}=\omega \frac{\left(1-\frac{v}{c}\right)}{\left(1+\frac{v}{c}\right)}
+\frac{\frac{2\pi m}{a} v}{1+\frac{v}{c}},
\end{equation}
\noindent and are identical to the the analytical result in~\cite{reed2005}. Note that under the substitution $m=0$, this relation reduces to the regular Doppler effect, where the reflected wave has been up-shifted to the frequency $\omega_{\text{ref},0}=\omega(1-v/c)/(1+v/c)$ due to the fact that the wall is approaching.

\section{Discussion}

We diagrammatically found that the negative harmonics experience an inverse Doppler shift. These harmonics all have positive phase velocity ($v_{ph}=\omega/k$) and negative group velocity ($v_{gr}=\D\omega/\D k$), and hence correspond to negative index media. This is consistent with the 1968 prediction of Veselago~\cite{veselago} that negative index media leads to inverse Doppler effect.

The reflected wave is a combination of the aforementioned harmonics. In order to obtain a global downshift effect, two strategies can be combined. The first would consist in minimizing the number of generated harmonics by substituting the moving wall to a smoother discontinuity. The second would be to design the PC so that the undesired $n=0$ space harmonic falls into a bandgap, as illustrated in Fig.~\ref{fig:ch5_dispersion}.

\section{Conclusion}

We have diagrammatically solved the problem of electromagnetic scattering for a wave normally incident on a photonic crystal bounded by a moving wall. This approach may be extended to the case of oblique incidence that will be also presented at the conference. Given its great simplicity and deep insight, the proposed diagrammatic approach may find wide applications in S-T electromagnetic systems.
\bibliographystyle{IEEEtran}
\bibliography{AP-S_2016}

\end{document}